\documentstyle[11pt]{article} 

\def\ds{\displaystyle}
\large{
\centerline{\bf Velocity quantization approach of the one-dimensional }
\vskip1pc
\centerline{\bf dissipative harmonic oscillator}
}
\centerline{G. L\'opez and  P. L\'opez}
\centerline{Departamento de F\'{\i}sica de la Universidad de Guadalajara}
\centerline{Apartado Postal 4-137}
\centerline{44410 Guadalajara, Jalisco, M\'exico}   
\centerline{PACS: 03.20.+i, 03.30.+p, 03.65.-w,03.65.Ca} 
\centerline{June, 2005}
\begin{document}  
\large{
\centerline{\bf ABSTRACT}
}
\vskip1cm
Given a constant of motion for the one-dimensional harmonic oscillator with linear dissipation in the velocity,
the problem to get the Hamiltonian for this system is pointed out, and the quantization up to second order in
the perturbation approach is used to determine the modification on the eigenvalues when dissipation is taken
into consideration. This quantization is realized using the constant of motion instead of the Hamiltonian.
\vfil\eject
\large{
\leftline{\bf 1. Introduction}    
}  
\vskip1pc\noindent
To find the Lagrangian and the Hamiltonian from the equations of motion of a given system (so
called "inverse problem of the calculus of variations" or "the inverse problem of the mechanics")
is not so trivial as one could think at first sight, even for one-dimensional systems and despite
their existence is guaranteed [1] here. If the system is autonomous (the forces do not depend
explicitly on time) or nonautonomous (total force depend explicitly on time), it is natural to
try to find Lagrangians and Hamiltonians which do no depend explicitly on time or which depend
explicitly on time for the latter. One possible approach to find these quantities for a
one-dimensional system is to find first a constant of motion of the system, and then to obtain
the Lagrangian and Hamiltonian [2]. This constant of motion, of course, is chosen to be
independent explicitly on time for autonomous systems, and explicitly depending on time, for
nonautonomous systems. It has been shown that when one tries to get quantities (constant of
motion, Lagrangian or Hamiltonian) which are explicitly depending on time for autonomous systems,
there may be a concern about ambiguities [3] or about consistence on the mathematical quantities
[4].
\vskip0.5pc\noindent
However, one common procedure that has been done is to guess an explicitly time
depending Hamiltonian for the harmonic oscillator with exponential time dependence of its mass
[5] which brings about the one-dimensional harmonic oscillator with linear dissipation in the
velocity. Indeed, there has been a lot of studies on the quantization of this Hamiltonian [6] and
decoherence of this quantum system [7]. However, since the one-dimensional harmonic oscillator
with linear dissipation on the velocity is an autonomous system, one should get time independent
dynamical quantities (constant of motion, Lagrangian or Hamiltonian) to describe it. After this,
one should proceed to make the quantization of this system in a consistent way. In fact, an explicitly
time independent constant of motion for this system has been found already [8]. So, in this paper 
we will study the classical damping behavior of the system with this constant of motion, and we
point out the difficulties to get its associated  Hamiltonian. Finally, we study its
quantization using the same constant of motion and the velocity operator. This last study is done
up to second order in perturbation theory and for weak dissipation.  
\vfil\eject
\leftline{\bf 2. Constant of motion and Lagrangian}
\vskip1pc\noindent
The one-dimensional harmonic oscillator with linear dissipation on the velocity is described by
the equation
$$m\ddot x+\alpha x+k x=0\ ,\eqno(1)$$
where $m$ is the mass of the particle, $\alpha$ is the dissipative constant, $k$ is the spring
constant, $x$ is the particle position, and $\dot x$ and $\ddot x$ are its first and second
differentiation with respect the time. By defining the new variable $\dot x=v$, Eq. (1) can be
written as the following autonomous dynamical system
$$\dot x=v\ ,\hskip2pc\hbox{and}\hskip2pc \dot v=-\omega^2x-2\omega_{\alpha}v\ ,\eqno(2)$$
where $\omega=\sqrt{k/m}$ is the natural angular frequency   of the harmonic oscillator without
dissipation, and $\omega_{\alpha}=\alpha/2m$ is the new dissipation parameter. A constant of
motion for the system (2) is a function $K=K(x,v)$ which satisfies the following partial
differential equation [9]
$$v~{\partial K\over\partial x}-\bigl(\omega^2x+
2\omega_{\alpha}v\bigr){\partial K\over\partial v} =0\ .\eqno(3)$$
The solution, $K_{\alpha}$, of this equation which satisfies the following  limit \break 
$\lim_{\alpha\to 0}K_{\alpha}=m v^2/2+m\omega_{\alpha}x^2/2$  is given by [8]
$$K_{\alpha}(x,v)={m\over 2}\biggl(v^2+2\omega_{\alpha}x v+\omega^2
x^2\biggr)~e^{-2\omega_{\alpha}G(v/x,\omega,\omega_{\alpha})}\ ,\eqno(4a)$$
where the function $G$ has been defined as
$$G(v/x,\omega,\omega_{\alpha})=\cases{{\ds 1\over\ds 2\sqrt{\omega_{\alpha}^2-\omega^2}}
\ln\left[{\ds\omega_{\alpha}+v/x-\sqrt{\omega_{\alpha}^2-\omega^2}\over\ds
\omega_{\alpha}+v/x+\sqrt{\omega_{\alpha}^2-\omega^2}}\right]&if~$\omega^2<\omega_{\alpha}^2$\cr\cr
{\ds 1\over\ds \omega_{\alpha}+v/x}&if~$\omega^2=\omega_{\alpha}^2$\cr\cr
{\ds 1\over\ds\sqrt{\omega^2-\omega_{\alpha}^2}}\arctan{\left({\ds\omega_{\alpha}+v/x\over\ds
\sqrt{\omega^2-\omega_{\alpha}^2}}\right)}&if~$\omega^2>\omega_{\alpha}^2$\cr}\eqno(4b)$$
For weak dissipation, one has the following expression for the constant of motion
$$K={1\over 2}mv^2+{1\over 2}m\omega^2 x^2+{m\omega_{\alpha}\over\omega}\biggl[
xv\omega-(v^2+\omega^2x^2)\arctan\left({v\over wx}\right)\biggr]\ .\eqno(5)$$
The Lagrangian for the system (1) can now be constructed from the known expression [2]
$$L(x,v)=v\int^v{K(x,\xi)~d\xi\over\xi^2}\ .\eqno(6)$$
For the general case (4), it is no possible to get a close expression for the Lagrangian, but for
the weak dissipation case (5), one gets 
$$L={1\over 2}mv^2-{1\over 2}m\omega^2x^2+{m\omega_{\alpha}\over\omega}\biggl[
(\omega^2x^2-v^2)\arctan{\left(v\over\omega x\right)}+\omega x v\ln\left({\omega^2x^2+v^2\over
\omega^2 x^2}\right)\biggr]\ .\eqno(7)$$
The generalized linear momentum ($p=\partial L/\partial v$) is 
$$p=mv+{m\omega_{\alpha}\over\omega}\biggl[\omega x+x\omega\ln\left({\omega^2x^2+v^2\over\omega^2
x^2}\right)-2v\arctan{\left({v\over\omega x}\right)}\biggr]\ .\eqno(8)$$
To get the Hamiltonian, it is necessary from (8) to express the variable $v$ as a function of the
variables $x$ and $p$, $v=v(x,p)$. In this way, one makes the substitution of this variable on the
Legendre transformation, $H(x,p)=v(x,p)~p-L(x,v(x,p))$, or in the constant of motion,
$H(x,p)=K(x,v(x,p))$. However, one notices immediately form (8) that it is not possible to do this. Thus,
the Hamiltonian can not be given explicitly but implicitly through the constant of motion (5).
\vskip1pc\noindent
A trajectory in the phase space ($x,v$) can be seen on Fig. 1, where the constant of motion (5)
has been used. Of course, in order to keep the continuity at ($v=0,x<0$), the arctan function
changes its value due to the multivalue functions. This means that our constant of motion
is really a local constant of motion (which is valid on the half plane $v<0$ or $v>0$), and it
changes its value every time the trajectory crosses the line $v=0$. This number of crossing is a
numerable set. Therefore, this set has measure zero [10]. In this way, one can say that (4) or
(5) represents a constant of motion almost everywhere in the phase space.
\vskip1pc\noindent
Let us now change the variables ($x,v$) by a new variables ($\phi,J$) defined as
$$\phi=\arctan\left({v\over\omega x}\right)\ ,\hskip2pc\hbox{and}\hskip2pc J={m\over
2\omega}\left(v^2+\omega^2x^2\right)\ ,\eqno(9a)$$
where the inverse transformation is given by
$$x=\sqrt{2J\over m\omega}~\cos\phi\ ,\hskip2pc\hbox{and}\hskip2pc v=\sqrt{2\omega J\over
m}~\sin\phi\ .\eqno(9b)$$
The dynamical system (2) is then transformed to the system
$$\dot\phi=-\biggl(\omega+{2\omega_{\alpha}\tan\phi\over 1+\tan^2\phi}\biggr)\eqno(10a)$$
and
$$\dot J=-4\omega_{\alpha}J\sin^2\phi\ .\eqno(10b)$$
These equations are readily solved, and their solutions are
$$\phi(t)=-\arctan\biggl[\tan(\omega t+a)-{\omega_{\alpha}\over\omega}\biggr]\eqno(11a)$$
and
$$J(t)=J_o~e^{-\omega_{\alpha}f(t)}\ ,\eqno(11b)$$
where $a$ and $J_o$ are constants determinate by the initial conditions, and $f(t)$ si defined as
$$f(t)={4\bigl(\tan{\omega t}-\omega_{\alpha}/\omega\bigr)^2t\over
1+\bigl(\tan{\omega t}-\omega_\alpha/\omega\bigr)^2}\ .\eqno(11c)$$
To know the trajectory in the new phase space ($\phi,J$), one can express $J$ as a function of
$\phi$ through the integration of  $\dot J=(dJ/d\phi)\dot\phi$. This brings about
the following expression
$$J(\phi)={\tilde
J_o\over\omega+\omega_{\alpha}\sin\phi}~e^{{2\omega_{\alpha}\over\omega}
\ds\arctan\bigl(\tan\phi+\omega_{\alpha}/\omega\bigr)}\ ,\eqno(12)$$
where $\tilde J_o$ is another constant. Fig. 2 shows Trajectories in this phase space for
$\omega_{\alpha}=0$ and for $\omega_{\alpha}\not=0$, where the expected jumps at $\pi/2$ and $3\pi/2$ are 
clearly seen at the scale shown.
\vskip1pc\noindent
To finish the classical analysis, let us
write the constant of motion, the Lagrangian and the generalized linear momentum for the weak
dissipative case in terms of the variables $\phi$ and $J$. These are given by
$$K=\omega J+\omega_{\alpha} J\biggl[\sin(2\phi)-2\phi\biggr]\ ,\eqno(13a)$$
$$L=-\omega
J\cos(2\phi)+2\omega_{\alpha}J\biggl[\cos(2\phi)-4\sin(2\phi)\ln(\cos\phi)\biggr]\eqno(13b)$$
and
$$p=\sqrt{2m\omega J}~\sin\phi+\omega_{\alpha}\sqrt{2mJ\over\omega}~\biggl[
\cos\phi\biggl(1-2\ln(\cos\phi)\biggr)-2\phi\sin\phi\biggr]\eqno(13c)$$
Once again, one see the impossibility to get the Hamiltonian $H(\phi,J)$ due to complexity of
the expression (13c).
\vskip3pc
\leftline{\bf 3. Quantization of the constant of motion}
\vskip1pc\noindent
Due to the impossibility of getting the Hamiltonian explicitly for the autonomous dynamical
system (2), one may propose to extend the Shr\"odinger quantization to a dissipative system
through the quantization of the constant of motion associated to it. This can be made by
associating an Hermitian operator to the velocity as
$$\widehat v=-{i\hbar\over m}{\partial\over\partial x}\ .\eqno(14)$$
In this way, if $\widehat K$ is the Hermitian operator associated to the constant of motion
(which must have units of energy), the associated Shr\"odinger equation of the classical
autonomous system would be 
$$i\hbar{\partial\Psi\over\partial t}=\widehat K(x,\widehat v)~\Psi\ ,\eqno(15)$$
where $\Psi=\Psi(x,t)$. Of course, the whole quantum mechanics structure is exactly the same
but with the velocity operator instead of the linear momentum operator as the main operator of
the quantum system. Since (15) represents an stationary problem, one just has to solve the
eigenvalue problem
$$\widehat K(x,\widehat v)~\phi=E\phi\ ,\eqno(16)$$
where $\phi=\phi(x)$ and one has used $\Psi(x,t)=\phi(x)\exp(-iEt/\hbar)$ in (15). If
$\widehat K$ can be written as $\widehat K=\widehat K_o+\widehat K_I$, where the solution of
the problem $\widehat K_o|n\rangle=E_n^{(0)}|n\rangle$ is known, the eigenvalues of (16) are given at first
order in perturbation theory [11] by
$$E_n=E_n^{(0)}+\langle n|\widehat K_I|n\rangle\ ,\eqno(17)$$    
where one has used Dirac notation [12]. Our constant of motion (5) or (13a) can be expressed of
the form $K=K_o+K_I$, where $K_o=\omega J$ and\break $K_I=\omega_{\alpha}J(\sin(2\phi)-2\phi)$.
It is well known that the eigenvalues of the harmonic oscillator without dissipation are given by
$E_n^{(0)}=\hbar\omega(n+1/2)$. Thus, $\widehat K_o$ is diagonal in the basis $\{|n\rangle\}$ and
can be expressed in terms of ascent , $a^+$, and descent, $a$, operators as
$$\widehat K_o=\hbar\omega(a^+a+1/2)\ ,\eqno(18)$$
with $a$ and $a^+$  defined in terms of $\widehat x$ and $\widehat v$ as
$$a=\sqrt{m\over 2\hbar\omega}~(\omega\widehat x+i\widehat v)\ ,\hskip1pc\hbox{and}\hskip1pc 
a^+=\sqrt{m\over 2\hbar\omega}~(\omega\widehat x-i\widehat v)\ .\eqno(19)$$
These operator have the following commutation relations
$$[a,a^+]=1\ , [a,a]=[a^+,a^+]=0\ , [a,a^+a]=a\ , [a^+,a^+a]=-a^+\ .$$
Additionally, $\widehat N=a^+a$ is called the number operator and is diagonal in the basis
$\{|n\rangle\}$, $\widehat N|n\rangle=n|n\rangle$, This, in turns, implies that the operator
associated to the variable $J$ is given by
$$\widehat J=\hbar(\widehat N+1/2)\ .\eqno(20)$$ 
Thus, our main problem is to assign a Hermitian operator to the function\break
$K_I=2\omega_{\alpha} J(\cos\phi\sin\phi-\phi)$. According to refence [13], one has the following
assignments
$$\sin\phi\longrightarrow\widehat S\ ,\cos\phi\longrightarrow\widehat C\ ,\phi\longrightarrow
\widehat\phi\ ,\eqno(21)$$
where $\widehat S$ and $\widehat C$ are Hermitian operators such that
$$\ddot{\widehat S}+\omega^2\widehat S=\ddot{\widehat C}+\omega^2\widehat C=0\ ,\hskip1pc
[\widehat C,\widehat N]=i\widehat S\ ,\hskip1pc [\widehat S,\widehat N]=-i\widehat C\eqno(22a)$$
and
$$\widehat S|n\rangle={i\over 2}\biggl(|n+1\rangle-|n-1\rangle\biggr)\ ,
\widehat C|n\rangle={1\over 2}\biggl(|n+1\rangle+|n-1\rangle\biggr)\ ,
[\widehat C,\widehat S]={\pi_o\over 2i}\ ,\eqno(22b)$$
with $\pi_o$ is the projector on the ground state of the quantum harmonic oscillator,
$\pi_o=|0\rangle\langle 0|$. Therefore, $\widehat C$ and $\widehat S$ commute for any exited
state. The operators $\widehat \phi$ is defined as
$$\widehat\phi={\pi\over 2}-\sum_{k=0}^{\infty}{(-1)^k\over 2k+1}\pmatrix{-1/2\cr k}\widehat
C^{2k+1}\eqno(23)$$
where $\pmatrix{a\cr b}$ si the combinatorial factor. In this way, the operator associated to the
function $K_I$ can be written as
$$\widehat K_I={\omega_{\alpha}\over 3}\biggl[\widehat J\widehat C\widehat S+
\widehat C\widehat J\widehat S+\widehat C\widehat S\widehat J+
\widehat J\widehat S\widehat C+\widehat S\widehat J\widehat C+
\widehat S\widehat C\widehat J\biggr]
-{\omega_{\alpha}}\biggl[\widehat J\widehat\phi+\widehat\phi\widehat J\biggr]\ .\eqno(24)$$
The first term has not contribution at first order in perturbation theory because of (22b), the
action of the operators $\widehat C$ or $\widehat S$ increases or decreases the state number such
that the expected value $\langle n|..|n\rangle$ is always zero. Due to the same reason, the 
contribution of the second term of
Eq. (24) will come only from the term $\pi/2$ of (23). Thus, at first order in perturbation theory,
one has the following correction of the eigenvalues
$$\delta E_n^{(1)}=-\hbar\omega_{\alpha}\pi(n+1/2)\ .\eqno(25a)$$
Thus, one gets
$$E_n\approx E_n^{(0)}+\delta E_n^{(1)}=
\hbar\omega(n+1/2)\biggl[1-{\omega_{\alpha}\pi\over\omega}\biggr]\ .\eqno(26a)$$ 
That is, there is a small shift on the frequency of oscillation given by
$$\omega'=\omega-\pi\omega_{\alpha}\ .\eqno(26b)$$
\vskip1pc\noindent
At second order in perturbation theory,$E_n=E_n^{(0)}+\delta E_n^{(1)}+\delta E_n^{(2)}$, the correction on the
energy will come from the expression
$$\delta E_n^{(2)}=\sum_{k\not=n}{|\langle k|\widehat K_I|n\rangle|^2\over E_n^{(0)}-E_k^{(0)}}
\ .\eqno(27)$$
Using (22b), (23) and (24), one has
\begin{eqnarray*}
\langle k|\widehat K_I|n\rangle&=& 
{\hbar\omega_{\alpha}\over 12}\biggl[(2k+4n+5)\delta_{k,n+2}-(2k+4n+1)\delta_{k,n-2}\biggr]\\
& &+\hbar\omega_{\alpha}(n+k+1)\sum_{l=0}^{\infty}\sum_{s=0}^{2l+1}{(-1)^l\over 2l+1}
\pmatrix{-1/2\cr l}\pmatrix{2l+1\cr s}\delta_{k,n-2l-1-s}
\end{eqnarray*}$$\eqno(28)$$
Therefore, the correction at second order can be written as
$$\delta E_n^{(2)}=
-{\hbar\omega_{\alpha}^2\over\omega}
\left[({2\over 3}n+{1\over 4})-\sum_{l=0}^{\infty}\sum_{s=0}^{2l+1}
\pmatrix{-1/2\cr l}^2\pmatrix{2l+1\cr s}^2{(2n-2l-s)^2\over (2l+1)^2(2l+1+s)}\right]\eqno(28)$$
\vfil\eject
\leftline{\bf Conclusions}
\vskip1pc\noindent
We have used the constant of motion for the one-dimensional dissipative harmonic oscillator to
study the classical trajectories in the phase space and to point out the difficulty to get its
Hamiltonian explicitly. Due to this problem, we have proposed the quantization of the constant
of motion directly , via the association of the velocity operator. Then, the eigenvalues where
calculated up to second order within perturbation theory to see the first effect of the dissipation on
them. This effect at first order corresponds to have a shift by the quantity $\pi\omega_{\alpha}$ on the
frequency of oscillation of the nondissipative harmonic oscillator. 
\vfil\eject
\leftline{Figure captions}
\vskip1pc\noindent
Fig. 1 Trajectory on the phase space ($x,v$) as determinate by (4), where $m=1~Kgr$, $\omega=1~sec^{-1}$ and
$\omega_{\alpha}=0.001~sec^{-1}$.
\vskip1pc\noindent
Fig. 2 Trajectory on the phase space ($\phi,J$) as determinate by (12). The straight horizontal line
corresponds to $\omega_{\alpha}=0$, solid line corresponds to \break $\omega_{\alpha}=0.001~sec^{-1}$,
 $\omega=1~sec^{-1}$ and $\tilde J_o=1$~Joules-sec.

\vfil\eject
\leftline{\bf References}
\vskip2pc\noindent
\obeylines{
1. D. Darboux,{\it Le\c cons sur la th\'eorie g\'en\'eral des surfaces et les applications 
\quad g\'eom\'etriques du calcul infinit\'esimal}, IVi\'eme partie,
\quad Gauthoer-Villars, Paris, 1984.
2. J.A. Kobussen, Act. Phys. Austr., {\bf 51} (1979) 193.
\quad C. Leubner, Phys. Rev. A {\bf 86} (1987) 9.
\quad C.C. Yan, Amer. J. Phys., {\bf 49} (1981) 296.
\quad G. L\'opez, Ann. Phys., {\bf 251}, 2(1996) 363.
3. G. L\'opez, Int. Jour. Theo. Phys., {\bf 37},5 (1998) 1617. 
4. G. L\'opez and J.I. Hern\'andez, Ann. of Phys., {\bf 193}, 1 (1989) 1.
5. P. Caldirola, Nuovo Cimento, {\bf 18} (1941) 393.
\quad E. Kanai, Prog. Theo. Phys., {\bf 3} (1948) 440.
\quad K.A. Yean and C.I. Um, Phys. Rev. A, {\bf 36} (1987) 5287.
6. V.V. Dodonov and M.S. Man'ko, Phys. Rev. D, {\bf 20} (1979) 550.
\quad R.K. Colegrave and M.S. Abdalla, J. Phys. A, {\bf 14} (1981) 2269.
\quad P.G.L. Leach, J. Phys. A, {\bf 16} (1983) 3261.
\quad H. Dekker, Phys. Rev. A, {\bf 16}, 5 (1977) 2126.
\quad H. Dekker, Phys. Rep. {\bf 80},1 (1981) 1-112.
7. M. Resenau da Costa, A.O. Caldeira, S.M. Deutra and H. Westfohl J.
\quad Phys. Rev. A, {\bf 61} (2000) 022107.
8. G. L\'opez, Ann. of Phys., {\bf 251},2 (1996) 372.
9. F. John, {\it Partial Differential Equations}, Springer-Verlag, N.Y. 1974.
\quad G. L\'opez, {\it Partial Differential Equations of First Order and Their
\quad Applications to Physics}, World Scientific, 1999.
10. E. Hewitt and K. Stromberg,{\it Real and Abstrac Analysis},
\quad Springer-Verlag N.Y., chapter III, 1965.
11. A. Messiah, {\it Quantum Mechanics}, vol.I, John Wiley and Sons, 1958.
12. P.A.M. Dirac, {\it The Principles of Quantum Mechanics}, IV edition,
\quad Oxford Science Publications, 1992.
}
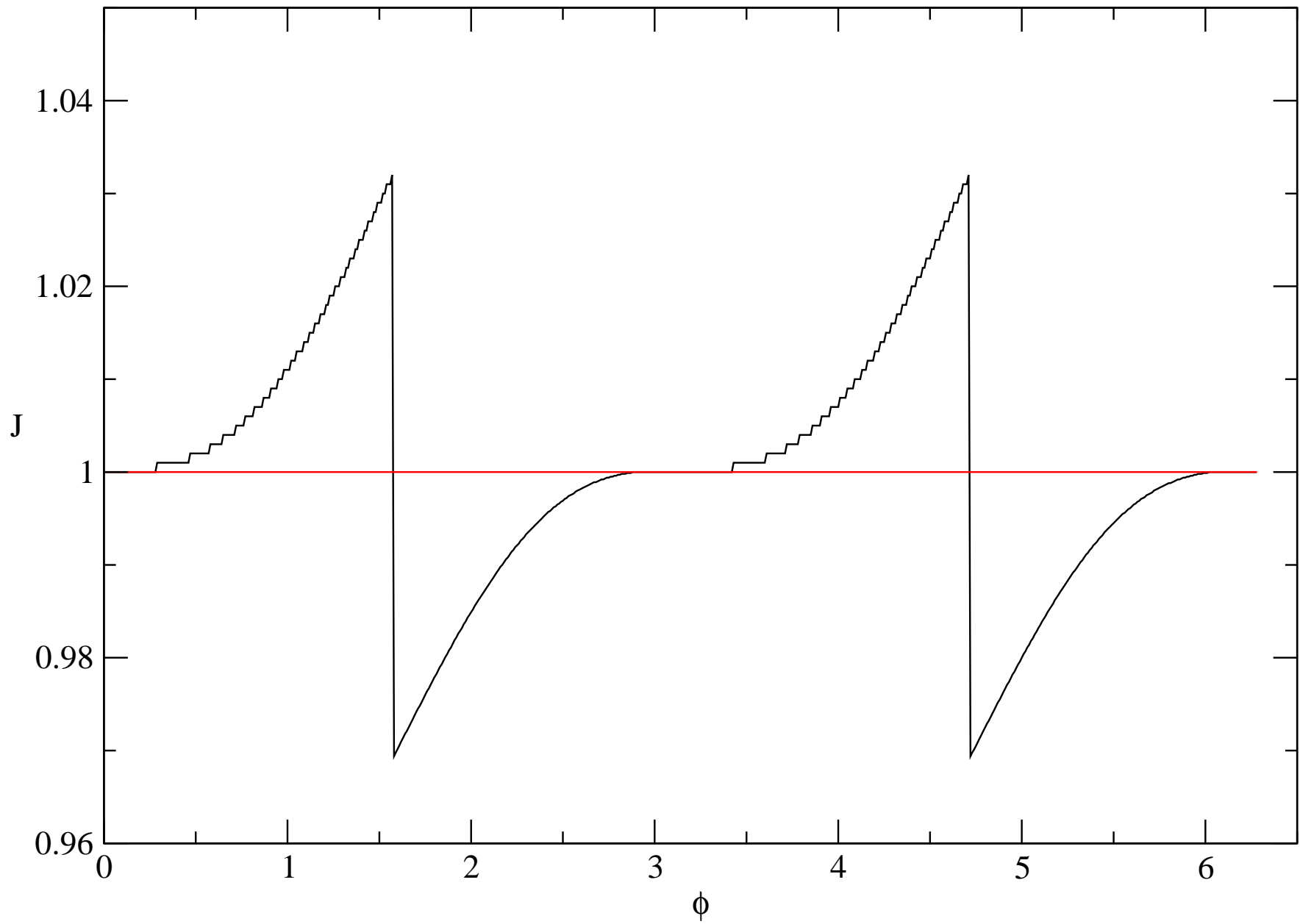
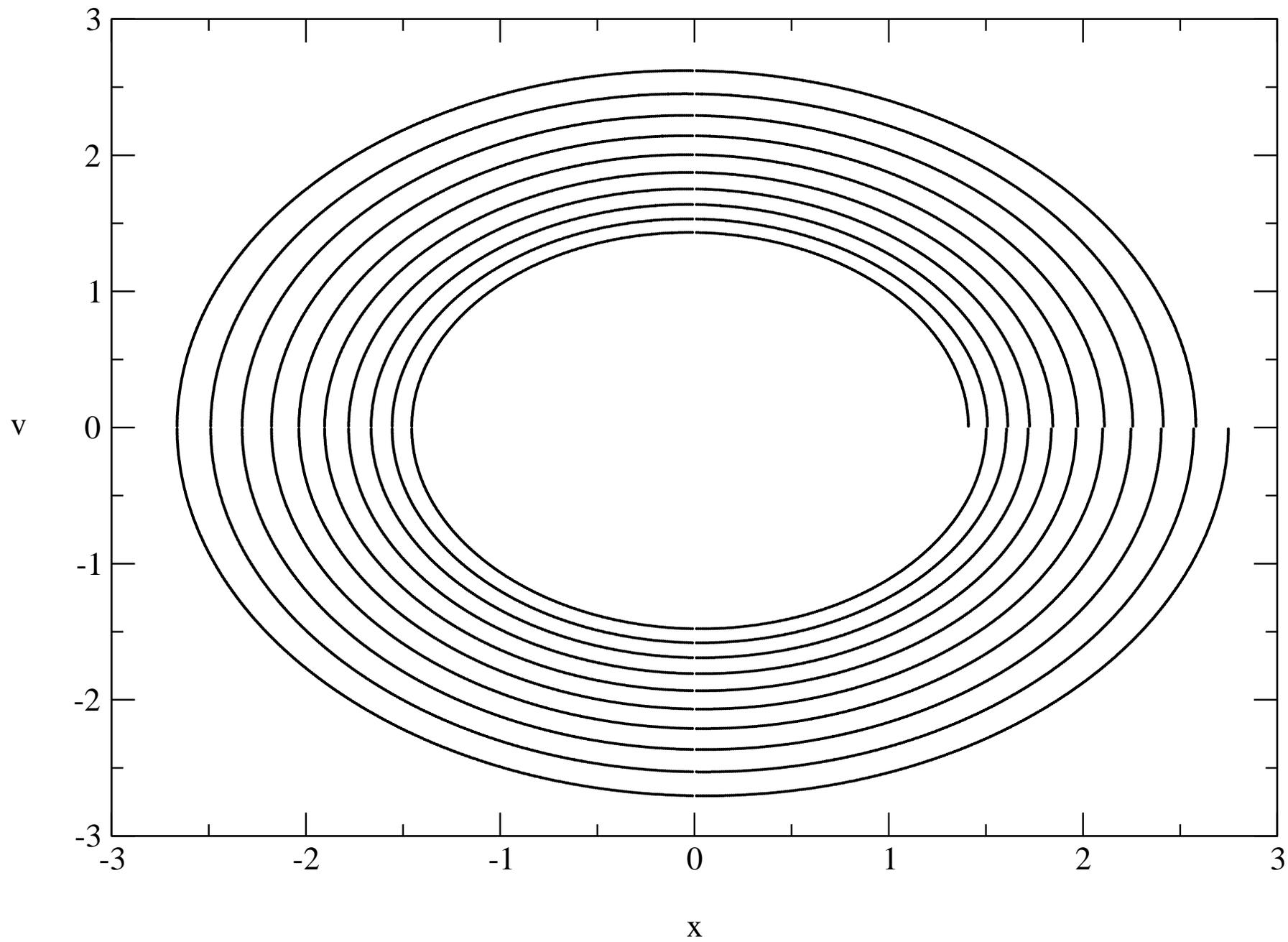
\end{document}